\documentclass[11pt,a4paper]{article}

\usepackage[utf8]{inputenc}
\usepackage[T1]{fontenc}
\usepackage[english]{babel}

\usepackage{amsmath}
\usepackage{amssymb}
\usepackage{graphicx}
\usepackage{booktabs}
\usepackage{siunitx}
\usepackage{geometry}
\usepackage{hyperref}
\usepackage{authblk}
\usepackage{booktabs}
\usepackage{tabularx}
\usepackage{amsmath}
\usepackage{multirow}
\title{Numerical Simulation of Electrical Properties in Cortical and Trabecular Bone: A Simplified Model}
\author[1]{María José Cervantes}
\author[1]{Catalina A. Cely-Ortíz}
\author[1,2]{C. Manuel Carlevaro}
\author[1,2,*]{Ramiro M. Irastorza}

\affil[1]{Instituto de Física de Líquidos y Sistemas Biológicos, CONICET-UNLP,
Av. 59 N$^\circ$ 789, 1900 La Plata, Argentina}

\affil[2]{Universidad Tecnológica Nacional, Facultad Regional La Plata,
Centro de Investigación en Mecánica Experimental y Computacional (CIMEC),
Av. 60 esq. 124, 1900 Berisso, Buenos Aires, Argentina}

\affil[*]{\textbf{Corresponding author:} Ramiro M. Irastorza,
Instituto de Física de Líquidos y Sistemas Biológicos, CONICET-UNLP,
Calle 59 N$^\circ$ 789, La Plata, Argentina.
E-mail: \texttt{rirastorza@iflysib.unlp.edu.ar}
\par
\textbf{ORCID:} Ramiro M. Irastorza: 0000-0002-6455-3574}

\date{}

\begin{document}

\maketitle

\begin{abstract}
The electrical properties of biological tissues depend on their composition and microstructure and determine their response to applied electric fields. In bone tissue, these properties are closely related to its composition, porosity, and microstructural organization. However, simplified models that link bone microstructure
(trabecular and cortical) to its effective electrical properties while accounting for the electrical anisotropy of
cortical bone remain scarce. In this work, a multiscale approach was developed, combining finite element simulations with Bruggeman-based homogenization models. Trabecular bone was modeled using micro-CT derived geometries, and its effective conductivity was analyzed as a function of BV/TV and free water content.
Cortical bone was represented by a simplified three-dimensional network of Haversian canals (axial = z) and Volkmann canals (transverse), incorporating their directional organization through an anisotropic
Bruggeman formulation. The models captured the conductivity range reported experimentally for trabecular bone (approximately 25–200 mS/m) and reproduced the anisotropic behavior observed in the cortical FEM
simulations, with higher conductivity in the axial direction. In the generated cortical geometries, 65.5\% of the
total porosity was associated with axial Haversian canals. At the macroscale, the tibia model with lower
trabecular and cortical BV/TV and reduced cortical thickness showed an approximately 38-39\% reduction in
impedance relative to the reference model. These results indicate that microstructural variations in bone tissue
are reflected in its effective electrical properties and in the macroscopic impedance of the model, providing a
physical and computational basis for future investigations into the sensitivity of electrical measurements to
changes in bone microstructure.

\end{abstract}


\section{Introduction}

Low-frequency electromagnetic field interactions with biological tissues are widely employed in biomedical applications. In bone tissue in particular, frequencies below 10 MHz have been used for diagnostic purposes \cite{KimelNaor2016,Katscher2017,Xu2025}, to promote healing processes \cite{Lin2019,Sun2025}, and for electrical stimulation \cite{Khalifeh2018,Wang2024}. Numerous experimental studies have investigated the dielectric properties of bone, and a comprehensive overview is provided in the review by Amin et al. \cite{Amin2019}. Despite the intrinsically hierarchical organization of bone, most experimental investigations assume it to be homogeneous for simplicity. Accordingly, its electrical properties are typically assumed as bulk (effective) values \cite{Kangasmaa2022}. In parallel, current exposure guidelines and safety standards—including the ICNIRP (2010, 2020) guidelines and IEEE Std C95.6-2002 and C95.1-2020—provide limited consideration of tissue heterogeneity and anisotropy across their respective frequency ranges (from 0 Hz to 300 GHz). To bridge this gap, Laakso et al. \cite{Laakso_2026} emphasize that incorporating advanced electromagnetic interaction models that account for tissue anisotropy is essential for dosimetric evaluations to accurately reflect real-world exposure scenarios. This consideration becomes particularly relevant at frequencies below 1 MHz, where the dielectric properties of biological tissues may themselves exhibit anisotropic behavior, further complicating their characterization and the accurate assessment of electromagnetic exposure.

Bone tissue is primarily composed of two distinct structural types: cortical (compact) bone and trabecular (cancellous) bone. For trabecular bone, imaging techniques such as micro-computed tomography (micro-CT) enable the investigation of the relationship between microstructure and dielectric properties \cite{Sierpowska2006}. Realistic geometrical models derived from trabecular architecture have facilitated the calibration of effective medium formulations, such as the Bruggeman model, for scaling  purposes and easily considering the tissue microstructure. These approaches allow the quantification of parameters such as bone volume fraction (BV/TV) and free water content \cite{Cervantes2023}.

In contrast, modeling cortical bone remains more challenging. The spatial resolution of micro-CT is generally insufficient to fully resolve the complexity of the intracortical canal network, including the Haversian and Volkmann systems. Experimental measurements demonstrate pronounced anisotropy, with conductivity and permittivity varying significantly along the axial, radial, and tangential directions \cite{Amin2019, Gao2016, DeMercato1991}. This anisotropic behavior is likely associated with the structural organization of the canal network. To address this limitation, a preliminary macroscopic in silico modeling framework was introduced by Cervantes et al. \cite{Cervantes2025}, enabling the controlled incorporation of biologically inspired statistical variability, including canal density and anisotropy, with good agreement as compared to experimental observations.

In this work, we propose mixing rules obtained by fitting our in silico models and extend our previous approach to simulate both trabecular and cortical bone, explicitly accounting for the anisotropy and the porosity. The main objective of this work is to provide simple mixing rules for simulating electrical problems at the macroscale.

\section{Materials and Methods}

\subsection{Model geometry}
\label{sec:Models_geometry_description}

\begin{figure}[h]
 \centerline{\includegraphics[width=.9\columnwidth,draft=false]{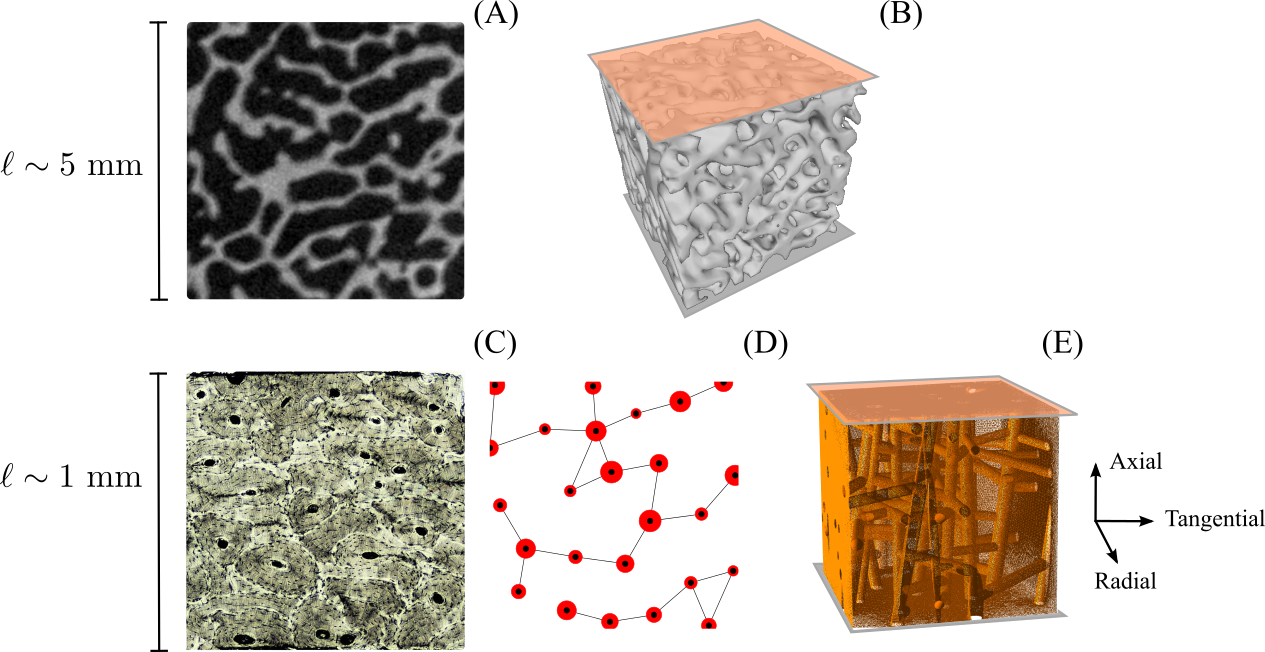}}
 \caption{A: MicroCT slice of a bovine trabecular bone. B: Cube generated from slices of real sample and modeling the three-dimensional geometry (see procedure in \cite{Cervantes2023}). C: Light micrograph 40x of human cortical bone (ground section, \url{https://commons.wikimedia.org/wiki/File:Connective_Tissue_Compact_Bone_(39978304590).jpg}). D: Haversian canals are shown in red (osteons not included in the model), and interconnections representing Volkmann canals in black. E: Three-dimensional geometry of the canal network within the cubic domain, showing the spatial organization and connectivity of the canals, as well as their orientation along the axial, tangential, and radial directions (see procedure in \cite{Cervantes2025}).}
 \label{fig:models}
\end{figure}

In this manuscript, two different numerical models were employed to estimate the effective conductivity of the two primary bone microstructures, namely trabecular and cortical bone.  The first, representing the trabecular bone, was described in \cite{Cervantes2023} (see Fig. \ref{fig:models} (A-B)). Briefly, the geometries were based on real samples of bovine trabecular bone acquired by micro-computed tomography (micro-CT).  Micro-CT scans were performed using a Bruker SkyScan 1173, and microstructural parameters (porosity) were computed with BoneJ. Bitmap images were processed to generate multiple geometrical models per sample. Two samples, A and B, were analyzed and several cubes were obtained per sample. Cubes (edge length 5 mm) were built and simulated using the finite element method (FEM). The size was chosen to analyze BV/TV effects while minimizing anisotropy influence \cite{Cervantes2023, Balmer2018}.

The second numerical model, representing cortical bone, consists of two primary components: the bone matrix and the intracortical canal network, composed of axial Haversian canals and transverse (communicating) Volkmann canals. The structural and geometrical parameters adopted in the model were based on histological data available in the literature (see \cite{RoblesLinares2019}, \cite{Cervantes2025} and the references therein). The model has five parameters (Table~\ref{tab:cortical_parameters}). The computational domain consisted of a 1 mm$^3$ cubic volume in which Haversian canals (axial direction) were distributed in a quasi-random arrangement that maximized spatial coverage while preventing non-physiological overlaps (see Fig. \ref{fig:models} (C-E)). This quasi-random, parameterizable approach enables the generation of multiple synthetic samples with controlled variability. Lacunae and canaliculi are not explicitly considered in the model; their contribution is assumed to be incorporated into the effective properties of the bone matrix, while only the larger Haversian and Volkmann canals are explicitly represented.

\begin{table}[ht]
\centering
\caption{Structural and geometric parameters of the synthetic cortical bone model.}
\label{tab:cortical_parameters}

\begin{tabular}{lccc}
\toprule
\textbf{Parameter} & \textbf{Range or Value} & \textbf{Unit} & \textbf{Comments} \\
\midrule

Osteon density 
& 10--25 
& \#/mm$^{2}$
& \\

Haversian canal diameter 
& 40--90 
& $\mu$m 
& \\

Haversian canal inclination angle 
& 0--15 
& Degrees 
& Relative to axial = z$^{*}$\\

Volkmann canal diameter 
& 40--90 
& $\mu$m 
&  \\

Volkmann canal inclination angle 
& 0--15 
& Degrees 
& Relative to Haversian canals$^{*}$\\

\bottomrule
\end{tabular}
\footnotesize
Data from Ref.~\cite{RoblesLinares2019}; $^{*}$ data from Ref.~\cite{Cervantes2025}.

\end{table}

\subsection{Numerical model}
\label{sec:Numerical_model}
Both models (trabecular and cortical) were based on the electric problem, which was solved numerically using FEM implemented with FEniCSx (version DOLFINx 0.9.0) \cite{Baratta2023}. The simulation scripts are publicly available in the repository \cite{soft_cervantes_2026}. As the biological medium can be considered almost fully resistive at 100 kHz \cite{Balmer2018}, the problem was approximated in its quasi-static scalar form. The electric potential $\varphi$ was computed from:
\begin{equation}
\nabla \cdot (\sigma \nabla \varphi) = 0
\label{eq:potential}
\end{equation}									                       	
where $\sigma$ is the electrical conductivity (S/m), which is a function of the position $(x,y,z)$. For the boundary conditions, active and passive electrodes were set at voltages  $\varphi_0$ and 0~V, respectively. A zero-current (homogeneous Neumann) condition was imposed on all the remaining outer surfaces. Then, the electric field was computed and the specific absorption rate (SAR) was integrated over the sample volume, computing the total power $P$. In order to obtain effective conductivity ($\sigma_{\mathrm{eff}}$), $P$ was set equal to the power dissipated in a lumped resistor of resistivity $\sigma_{\mathrm{eff}}^{-1}$, length $\ell$ and cross-sectional area $A$. This reasoning can be extended to tensorial quantities to properly account for the anisotropic nature of cortical bone. 

To guarantee numerical stability and independence of the finite element solutions, a systematic convergence and sensitivity analysis was conducted across all computational domains. For both the 5 $\text{mm}^3$ trabecular bone volumes and the 1 $\text{mm}^3$ cortical bone microstructural models, unstructured tetrahedral elements were generated, and mesh density was iteratively refined. The scalar quasi-static electric potential equation was solved using FEniCSx across approximately 1,500 simulation runs to evaluate the sensitivity of the effective electrical conductivity ($\sigma_{\text{eff}}$) to mesh resolution, electrical properties, and structural parameters. A convergence criterion was defined such that further element refinement produced less than a 1\% relative change in the computed effective conductivity tensor components ($\sigma_{\text{eff}}$). For the macroscale bovine tibia model, local mesh refinement was implemented at the contact interface between the 5 mm radius surface electrodes and the cortical domain, where spatial potential gradients are highest. This mesh configuration ensured stable and reproducible macroscopic impedance calculations ($Z_H$ and $Z_V$) under both healthy and non-healthy tissue conditions.

\subsection{Mixing-rules}
\label{sec:Mixing_rules}
In electromagnetism several mixing-rules exist \cite{Sihvola1999} and, in general, they are expressed as a function of the relative permittivity $\varepsilon$, which is a complex number that varies with frequency
(among other parameters). The conductivity enters the imaginary part of the complex relative permittivity,
\mbox{$\varepsilon = \varepsilon' - \frac{\sigma}{j\omega\varepsilon_0}$},
where $\omega$ is the angular frequency and $\varepsilon_0$ is the vacuum permittivity. All the formulations presented next will be expressed using the relative permittivity.

The Bruggeman model was already proposed for the case of the trabecular bone \cite{Cervantes2023}. In such a case, the anisotropy is not considered and it was validated for samples up to 5 mm in edge length. The effective permittivity is computed via: 
\begin{equation}
\sum_{i=1}^{N} f_i
\frac{\varepsilon_i-\varepsilon_{\mathrm{eff}}}
{\varepsilon_i+2\varepsilon_{\mathrm{eff}}}
= 0
\label{eq:bruggeman}
\end{equation}
where $N$ is the number of constituent media, while $f_i$ and
$\varepsilon_i$ represent the volume fraction and permittivity of
medium $i$, respectively. Unlike the Maxwell--Garnett model, which
assumes a host matrix with dispersed inclusions, the Bruggeman model
treats all constituent phases symmetrically \cite{Sihvola1999}. For trabecular bone,
three constituent media were considered: free water (or physiological
solution), bone marrow, and bone matrix \cite{Cervantes2023}.

On the other hand, the anisotropy of the cortical bone, along with its correlation to microstructure and mechanical behavior, is a well-established area of study \cite{Hoc2006}, and computational homogenization models for mechanical properties are widely developed \cite{Hollister1994}. However, there are few electrical homogenization models for cortical bone, and most do not account for its orthotropic anisotropy \cite{Gao2016,Ciuchi2013}. Here we propose the Bruggeman model with three inclusion families — axial, tangential and radial canals — embedded in a homogeneous bone matrix (four media in total). All inclusions are assumed to be isotropic, while the resulting anisotropy arises from their preferential orientation within the canal network. The implicit formulation can be written as follows \cite{Sihvola1999,Mackay2010,Mejdoubi2006}:
\begin{equation}
\sum_{i=1}^{4}
f_i (\varepsilon_i-\varepsilon_{\mathrm{eff}})
\cdot
\left[
I + N_i \cdot
(\varepsilon_i-\varepsilon_{\mathrm{eff}})
\cdot \varepsilon_{\mathrm{eff}}^{-1}
\right]^{-1}
= 0
\label{eq:anisotropic_bruggeman}
\end{equation}
Again, $f_i$ and $\varepsilon_i$ represent the volume fraction and
permittivity of medium $i$, respectively. $I$ denotes the identity
tensor, and $N_i$ is the depolarization tensor associated with the
$i$-th inclusion. It should be emphasized that all quantities in the
formulation are tensorial. The anisotropy of the model is encapsulated
in the depolarization tensor; for instance, for cylindrical inclusions
aligned with the $z$-axis, the tensor is given by:
\begin{equation}
N_i =
\begin{pmatrix}
\frac{1}{2} & 0 & 0 \\
0 & \frac{1}{2} & 0 \\
0 & 0 & 0
\end{pmatrix}.
\end{equation}
Simulations on three-dimensional models of finite length $H$ and radius $R$ showed that with ratio $H/R$ approximately 10 the component $N_i(3,3)$ is negligible \cite{Mejdoubi2006}. In the case of Haversian canals $H/R$ is around 30 (with samples of size of Fig.~\ref{fig:models}).

\subsection{Material properties}
\label{sec:Material_properties}
In previous works, we reviewed in detail the conductivity of bovine trabecular bone at 100 kHz \cite{Cervantes2023}. This frequency was selected because bone behaves as an almost purely resistive medium. The two tissues considered for numerical simulation, bone marrow and bone matrix, have respectively the ranges: 20--700~mS/m, and 0.2--21~mS/m \cite{Cervantes2023}. Fitting the FEM model described in section~\ref{sec:Numerical_model} yielded the following values: 300.0 mS/m and 20.1 mS/m. Regarding the free water content, the existing literature is limited; here we used data for human trabecular bone and free water content presented by Sierpowska et al. \cite{Sierpowska2007}, but the results should be carefully interpreted because the BV/TV is lower than that of bovine sample.

For cortical bone electrical properties and its relationship with microstructure, a revision of literature was done in \cite{Cervantes2025}. We concluded that little is known about the porosity and electrical conductivity of cortical bone. Gao et al. \cite{Gao2016} evaluated porosity as the ratio of liquid volume to total volume and reported a linear relationship between porosity and electrical conductivity. However, the frequency at which the conductivity measurements were performed was not specified. Regarding the anisotropy, in bovine cortical bone measured at 100 kHz, the conductivity in the axial direction (7.5 mS/m) was approximately 36\% higher than in the circumferential direction (5.52 mS/m) and 70\% higher than in the radial direction (4.4 mS/m) (values read from the curves in De Mercato and García Sánchez \cite{DeMercato1991}).  In humans, this difference was even more pronounced at 10 kHz, almost an order of magnitude \cite{Amin2019}. With this information at hand, we fitted the FEM numerical model and it yielded 3.8 mS/m for the hydrated cortical bone matrix and 267 mS/m for the conductivity of the porous domain (canal network) \cite{Cervantes2025}. 
Table~\ref{tab:tissue_properties} summarizes all the electrical and porosity parameters used in this manuscript to fit and to simulate the numerical models presented in  sections~\ref{sec:Models_geometry_description} and~\ref{sec:Numerical_model}.

\begin{table}[h]
\centering
\caption{Tissue properties of numerical FEM models. Values of conductivity were fitted to experimental results reported in \cite{Gao2016,Balmer2018}.}
\label{tab:tissue_properties}

\begin{tabular}{lcc}
\toprule
\textbf{Tissue / material} &
\textbf{Conductivity (mS/m)} &
\textbf{BV/TV = 1 - porosity} \\
\midrule

Trabecular bone matrix
& 20.1
& \multirow{2}{*}{0.285--0.724$^{*}$} \\

Trabecular bone porous
& 300.0
& \\

\midrule

Cortical bone matrix
& 3.8
& \multirow{2}{*}{0.90--0.95$^{**}$} \\

Cortical canal network
& 267.0
& \\

\bottomrule
\end{tabular}

\vspace{2mm}

\footnotesize
$^{*}$ From Refs.~\cite{Cervantes2023,Balmer2018};
$^{**}$ From Ref.~\cite{Gao2016}.

\end{table}

\section{Results}

\subsection{Trabecular bone}
\label{sec:Trabecular_bone}

\begin{figure}[h]
 \centerline{\includegraphics[width=1\columnwidth,draft=false]{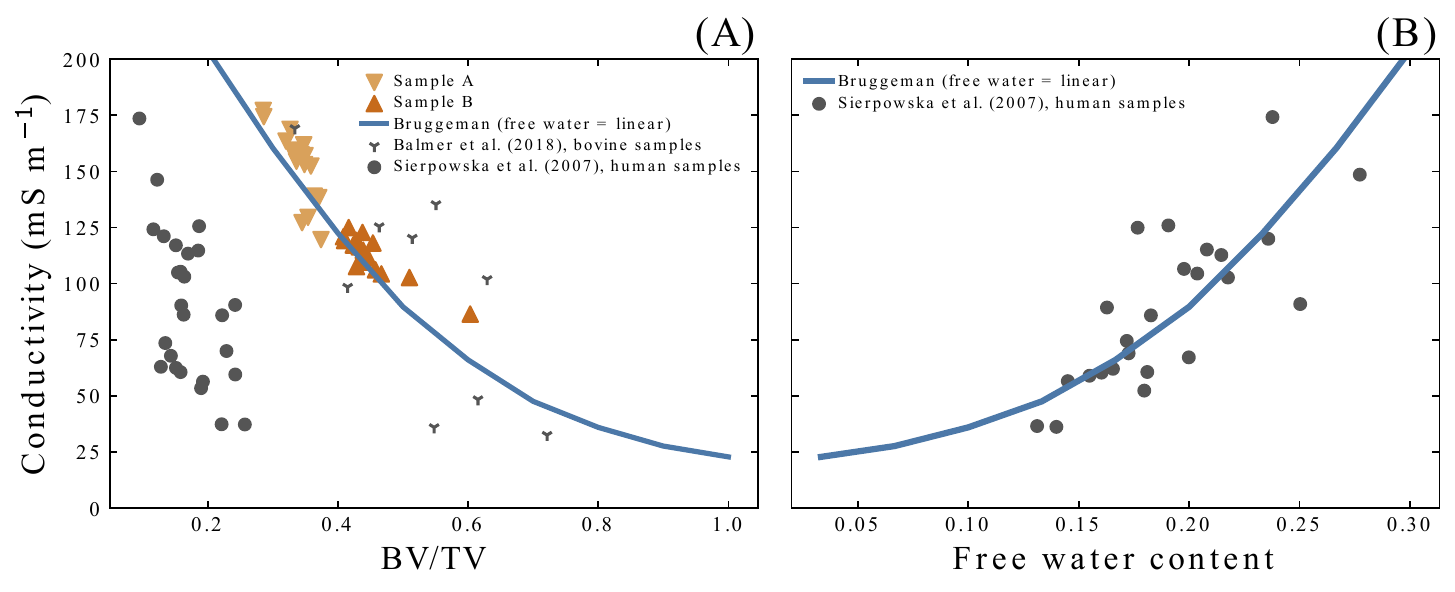}}
 \caption{Numerical simulations (samples A and B, see \cite{Cervantes2023}) and effective mixture rules obtained by Eq. \ref{eq:bruggeman}. A: Effective conductivity as a function of BV/TV ratio for bovine and human bone. B: Effective conductivity of human samples as a function of free water content. Continuous lines show the results for homogenization formulation of Bruggeman using the described procedure.}
 \label{fig:trabecular}
\end{figure}

Figure \ref{fig:trabecular} shows the effective conductivity predicted by the simulation and Bruggeman models, together with the experimental data reported by by Balmer et al. \cite{Balmer2018} (bovine) and Sierpowska et al. \cite{Sierpowska2006,Sierpowska2007} (human), as a function of BV/TV and free water content. Since, as far as we know, no quantitative data are available relating the free water content (with information of BV/TV) to the electrical conductivity of bovine samples, the effective conductivity predicted by the Bruggeman model was compared only with the human experimental data. The conductivity values ranged from approximately 25 mS/m to 200 mS/m for both bovine and human samples. Although, as previously reported, the BV/TV values differ substantially between the two species \cite{Sierpowska2006,Balmer2018}, the Bruggeman model accurately captures the conductivity behavior of bovine bone. The model also reproduces the observed dependence on free water content of human samples. However, this relationship should be interpreted with caution, as changes in free water content are intrinsically coupled with changes in BV/TV, both of which influence the effective electrical conductivity. The resulting Bruggeman parameters are summarized in Table~\ref{tab:bruggeman_parameters}.

\begin{table}[h]
\centering
\caption{Mixing-rule parameters for the Bruggeman model of cortical and trabecular bone at 100 kHz. See Eqs.~(\ref{eq:bruggeman})--(\ref{eq:anisotropic_bruggeman}).}
\label{tab:bruggeman_parameters}

\small
\renewcommand{\arraystretch}{1.35}

\begin{tabularx}{\textwidth}
{l l c c l >{\centering\arraybackslash}p{3.1cm}}

\toprule
\textbf{Bone} &
\textbf{Medium} &
\textbf{Conductivity} &
\textbf{Relative} &
$\mathbf{N}_i$ &
$f_i$ \\

&
&
\textbf{(mS/m)} &
\textbf{permittivity} &
&
\\
\midrule

Trabecular
& Matrix
& 20.1
& $2.28\times10^{2}$
& --
& $f_t$ \\

& Bone marrow
& 103.0
& $1.73\times10^{2}$
& --
& $\tfrac{19}{30}-\tfrac{2}{3}f_{t}$ \\

& Physiological solution
& 1200.0
& 78.0
& --
& $\tfrac{11}{30}-\tfrac{f_{t}}{3}$ \\

\midrule

Cortical
& Cortical matrix
& 3.8
& $2.28\times10^{2}$
& $\left(\begin{smallmatrix}
1/3&0&0\\
0&1/3&0\\
0&0&1/3
\end{smallmatrix}\right)$
& $f_c$ \\

& Axial canals
& 267.0
& $1.73\times10^{2}$
& $\left(\begin{smallmatrix}
1/2&0&0\\
0&1/2&0\\
0&0&0
\end{smallmatrix}\right)$
& $(1-f_c)p_v$ \\

& Tangential canals
& 267.0
& $1.73\times10^{2}$
& $\left(\begin{smallmatrix}
0&0&0\\
0&1/2&0\\
0&0&1/2
\end{smallmatrix}\right)$
& $\tfrac{(1-f_c)(1-p_v)}{2}$ \\

& Radial canals
& 267.0
& $1.73\times10^{2}$
& $\left(\begin{smallmatrix}
1/2&0&0\\
0&0&0\\
0&0&1/2
\end{smallmatrix}\right)$
& $\tfrac{(1-f_c)(1-p_v)}{2}$ \\

\bottomrule
\end{tabularx}

\vspace{2mm}

\footnotesize
$f_t$: trabecular BV/TV, range (0.10--0.8);\\
$f_c$: cortical BV/TV, range (0.85--1.00);\\
$p_v$: proportion of axial = z canals relative to total porosity.

\end{table}
\subsection{Cortical bone}
\label{sec:Cortical_bone}

\begin{figure}[h]
 \centerline{\includegraphics[width=1\columnwidth,draft=false]{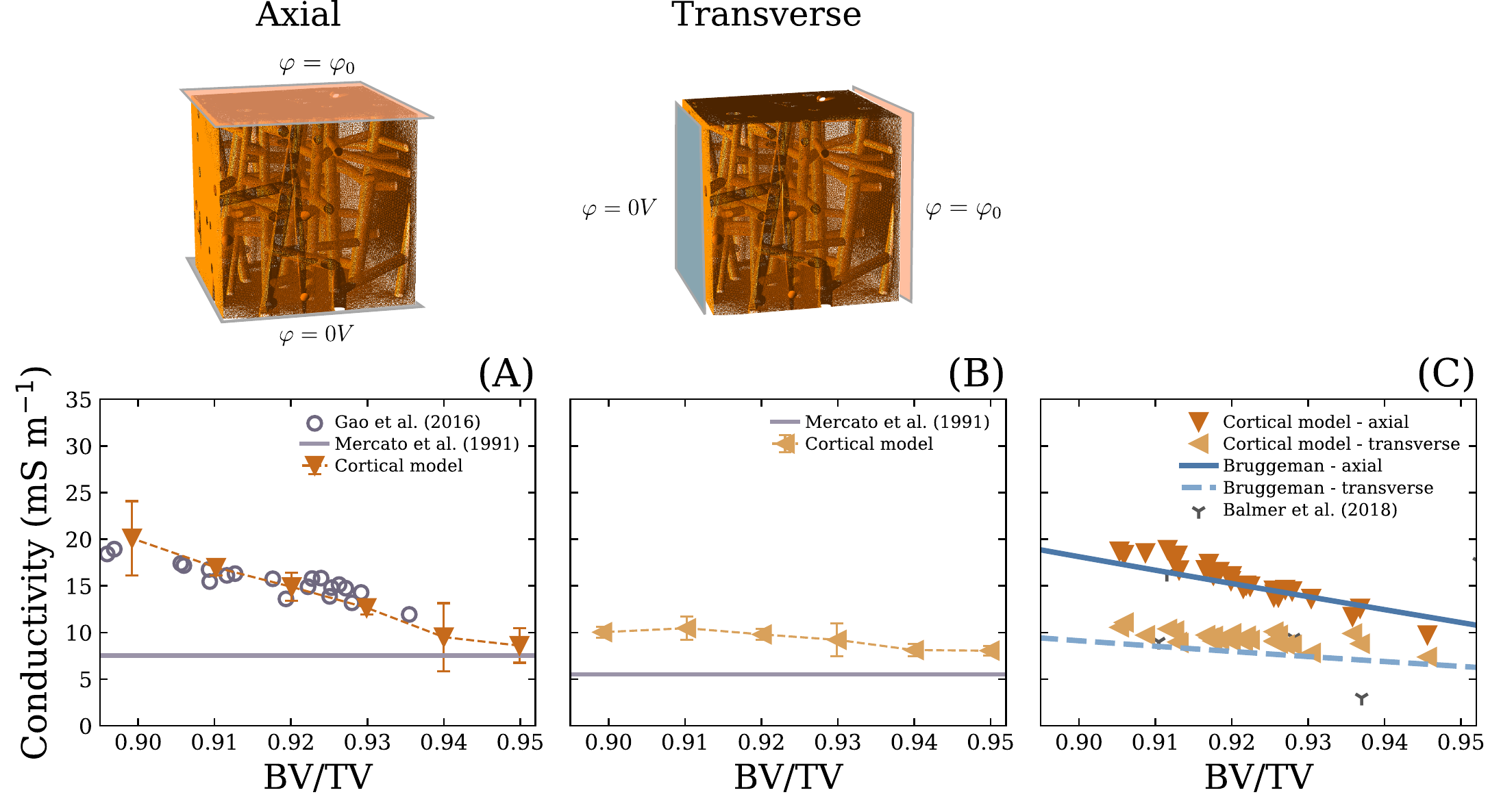}}
 \caption{Estimation of the effective conductivity of the cortical FEM model (numerical simulation). Axial (A) and transverse (B) effective conductivities of the samples. Error bars are the standard deviation obtained from multiple simulations with randomly oriented and distributed Haversian and Volkmann canals. Experimental data from \cite{Gao2016} are shown as circles, whereas those from \cite{DeMercato1991} are represented by the gray solid line (since the BV/TV values were not reported by the authors, the corresponding range is shown across the entire plotted BV/TV interval). (C) Numerical simulations and effective mixture rules obtained by Eq. \ref{eq:anisotropic_bruggeman} for several cortical bone models with different BV/TV values in the axial and transverse directions.}
 \label{fig:cortical1}
\end{figure}

Figure \ref{fig:cortical1} shows the result of the microscopic cortical FEM model together with the experimental data from De Mercato and García Sánchez \cite{DeMercato1991} and Gao and Sevostianov \cite{Gao2016}. The effective conductivity is shown as a function of BV/TV for the axial (Fig. \ref{fig:cortical1}A) and transverse (Fig. \ref{fig:cortical1}B) directions. Simulations in the axial and transverse directions were performed by changing the position of the two parallel square electrodes. 

Figure \ref{fig:cortical1}C shows the effective conductivity obtained from the cortical FEM simulations together with the predictions of the anisotropic Bruggeman model given by Eq. \ref{eq:anisotropic_bruggeman}, for both the axial and transverse directions. The relative distribution of the canal porosity was estimated from the cortical FEM geometries. The mean total porosity was 7.60\%, of which 4.98\% corresponded to the axial canal network. Accordingly, 65.5\% of the total porosity was assigned to axial Haversian canals, while the remaining 34.5\% was equally distributed between transverse canals oriented along the $x$ and $y$ directions. These relative fractions were kept constant while BV/TV was varied in the Bruggeman model. The anisotropic Bruggeman model reproduced the directional behavior observed in the FEM simulations, with higher effective conductivity in the axial direction than in the transverse direction. 

\subsection{Macroscale model}
\label{sec:Macroscale_model}
\begin{figure}[h]
 \centerline{\includegraphics[width=1\columnwidth,draft=false]{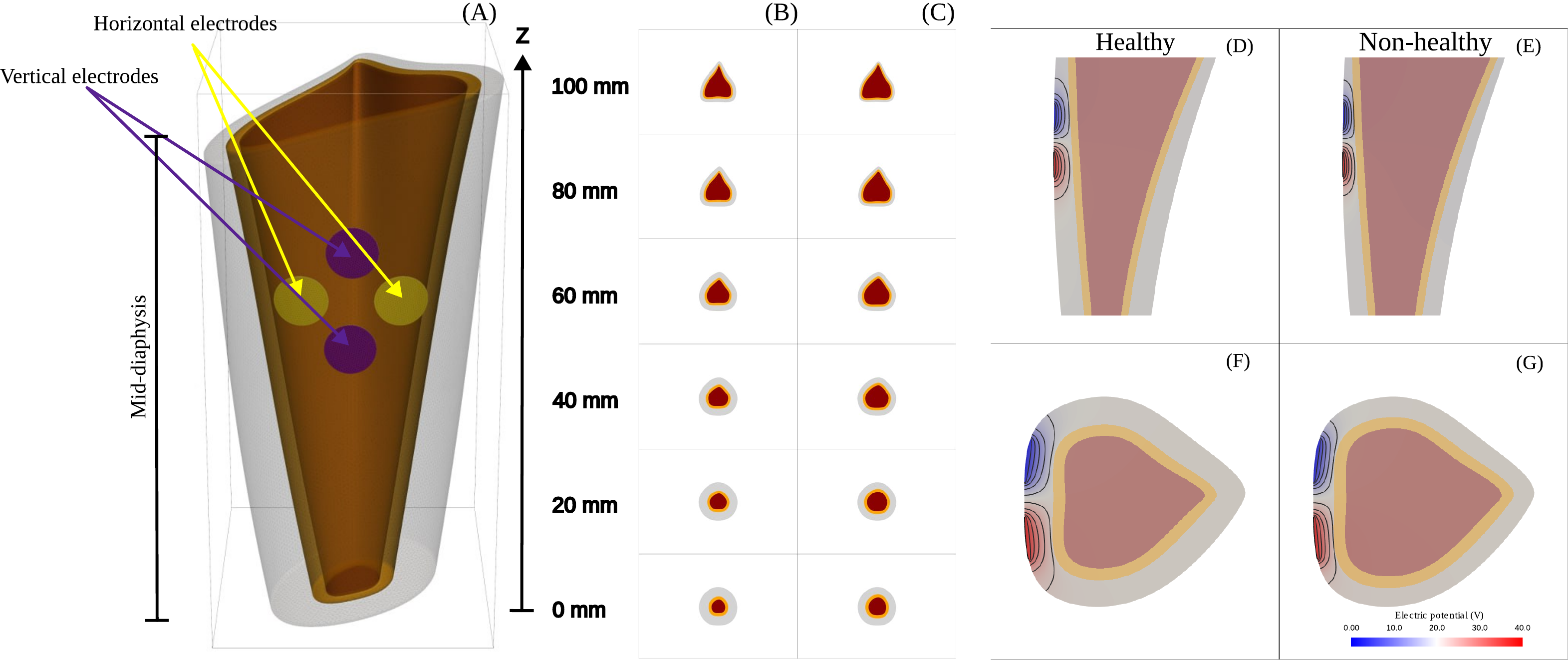}}
 \caption{Macroscale model of bovine tibia. A: Mid-diaphysis with the disposition of the electrodes to evaluate the anisotropy. Slices of the healthy (B) and non-healthy (C) tibia (gray cortical bone, yellow trabecular and red bone marrow). Electric potential distribution obtained from the macroscale FEM model for the healthy and non-healthy conditions. (D–E) axial views and (F–G) transverse views.}
 \label{fig:macro}
\end{figure}

As an application example, the effective electrical properties obtained from the mixture models were incorporated into a macroscale FEM model of a bovine tibia. The geometry was conveniently selected such that the principal material directions coincide with the local $x$-, $y$-, and $z$-axes, thereby simplifying the definition of the orthotropic electrical conductivity tensor (otherwise, the Appendix A provides details on how to account for the general case). The geometry was constructed using the proximal half of the bovine tibial diaphysis reported by Choi et al. ~\cite{Choi2006} as an anatomical reference. The model consisted of three domains representing cortical bone, trabecular bone, and bone marrow. Healthy and non-healthy conditions were considered by modifying the BV/TV and the corresponding effective electrical properties of the cortical and trabecular regions, as well as the cortical thickness. For the trabecular region, BV/TV values of 0.40 and 0.18 were selected based on the experimental porosity range reported for bovine cancellous bone by Karki and Wu \cite{Karki2022}. 
For the cortical region, BV/TV values of 0.95 and 0.90 were considered as representative conditions with different cortical porosity. The corresponding effective conductivities were calculated using the anisotropic Bruggeman model described in Section~\ref{sec:Mixing_rules}, with the axial direction of the Haversian canals defined as the $z$-axis. In addition, cortical thickness was reduced in the non-healthy model to represent the cortical thinning associated with bone deterioration reported by Zhu et al.~\cite{Zhu2016}. A reduction of 2~mm was considered in the present model. 
Figure \ref{fig:macro} shows the macroscale FEM model of the bovine tibia. Circular electrodes with a radius of 5~mm, comparable in size to commercially available surface electrodes, were used. The electrode configurations used to evaluate the impedance in the horizontal ($Z_{H}$) and vertical directions ($Z_{V}$) are shown in Fig. \ref{fig:macro}A, while cross-sections of the healthy and non-healthy models are shown in Figs. \ref{fig:macro}B and \ref{fig:macro}C, respectively.
The electric potential distribution was evaluated for both healthy and non-healthy models. Figures \ref{fig:macro}D–G show the potential distribution obtained from the macroscale FEM simulations for the two conditions. 
The impedance values obtained for the healthy and non-healthy models are summarized in Table \ref{tab:impedance_results} for the horizontal and vertical electrode configurations. 
The non-healthy model showed lower impedance values than the healthy model for both electrode configurations, with a reduction of approximately 39\%. Differences between the horizontal and vertical configurations were 2.0\% and 0.9\% for the healthy and non-healthy models, respectively. 
\begin{table}[ht]
\centering
\caption{Impedance results for healthy and non-healthy bone conditions.}
\label{tab:impedance_results}

\begin{tabular}{lccc}
\toprule
\textbf{Condition} &
$\mathbf{Z_H}$ \textbf{(k$\Omega$)} &
$\mathbf{Z_V}$ \textbf{(k$\Omega$)} &
$\mathbf{\Delta Z_{H-V}}$ \textbf{(\%)} \\
\midrule

Healthy
& 9.67
& 9.48
& 2.0 \\

Non-healthy
& 5.91
& 5.86
& 0.9 \\

\bottomrule
\end{tabular}

\end{table}

\section{Discussion}
\label{sec:Discussion}

The multiscale modeling approach presented in this work links the microscopic structure of bone tissue to its macroscopic electrical response. Microscale finite element models were developed using micro-CT-derived geometries for trabecular bone and simplified, literature-informed models for cortical bone, calibrated against available experimental data. The resulting simulations were used to derive Bruggeman mixing rules for the macroscopic tibia model.\\

At the tissue level, the Bruggeman model accurately captured the effective conductivity of both bone types. For trabecular bone, it successfully represented the bovine experimental data within a range of 25–200~mS/m, highlighting the critical, coupled roles of BV/TV and free water content. Although the model reproduces the dependence on free water, this behavior was necessarily compared with human data due to the absence of bovine experimental records for free water, BV/TV, and conductivity simultaneously. In physiological scenarios, fluid substitution and trabecular thinning are inherently coupled. Therefore, conductivity changes reflect a synchronized alteration of the matrix-fluid ratio rather than a single isolated structural parameter.\\

For cortical bone, the microscopic FEM and anisotropic Bruggeman models effectively captured the tissue's directional behavior. The higher effective conductivity observed in the axial direction compared to the transverse direction is a direct consequence of the spatial distribution of the canal network. Our geometrical analysis revealed that 65.5\% of the total porosity belongs to axial Haversian canals. This fluid-filled, highly conductive pathway aligned with the axial axis naturally facilitates current flow. The fact that the analytical Bruggeman model closely matched the FEM simulations is consistent with the parameter choice ($N_i$ tensors) of Table~\ref{tab:bruggeman_parameters} as a reliable tool for predicting cortical anisotropy without the need for heavy computational imaging meshes.\\

Expressions for monitoring bone health were proposed by Kimel-Naor et al. \cite{KimelNaor2016}, in which electrical conductivity was computed as a function of bone mineral density, distinguishing between normal bone, mild-to-moderate-to-advanced osteopenia, and osteoporosis. Applying these relationships to the cortical and cancellous bone densities considered in this study, resulted in conductivity ranges of 12.48--38.73 mS/m for cortical bone and 58--229 mS/m for cancellous bone. These values are in good agreement with the conductivity ranges predicted by the models developed in the present study, supporting the consistency of the proposed microscale modeling. In addition, unlike the relationships proposed by Kimel-Naor et al., the present model accounts for the anisotropic behavior of bone tissue, providing a more comprehensive representation of its electrical properties.\\

When scaling up to the macroscale bovine tibia model, the incorporation of these effective properties yielded a significant 38-39\% reduction in total impedance for the non-healthy bone condition across both electrode configurations. This marked drop in impedance ($Z_H$ and $Z_V$) is driven by the combined effect of three factors modeled simultaneously:
\begin{enumerate}
    \item The reduction of trabecular BV/TV (from 0.40 to 0.18), which increases the volume fraction of the highly conductive physiological solution (1200~mS/m).
    \item The increase in cortical porosity (BV/TV drop from 0.95 to 0.90), increasing the presence of conductive fluid canals.
    \item The 2~mm reduction in cortical thickness, which shortens the distance that the current must travel through the most resistive layer of the bone.
\end{enumerate}

Interestingly, electrode orientation showed very low sensitivity; the difference ($\Delta Z_{H-V}$) dropped from 2.0\% in healthy bone to 0.9\% in non-healthy bone. This suggests that macroscale structural degradation (cortical thinning and trabecular loss) homogenizes the current path, overriding microscopic anisotropy. These in silico results suggest that although the effective electrical impedance measured at the surface is highly sensitive to overall bone degradation, detecting localized or direction-dependent damage solely through the rotation of this type of surface electrode remains challenging.
\subsection{Limitations}

When interpreting these results, some limitations of the present study should be considered. First, the model assumes that changes in BV/TV are primarily reflected in variations in the relative fractions of the bone and fluid phases, without explicitly representing the complex microstructural changes associated with bone remodeling. In particular, changes in pore size, shape, connectivity, and spatial distribution that may occur in osteoporosis are not explicitly resolved. Second, the macroscale model considered simplified isotropic domains for bone marrow and idealized electrode geometries. Third, this study addressed only electrical conductivity and did not account for the contribution of permittivity to the electrical response of bone tissue. Fourth, although the tissue-scale models were compared with experimental data available in the literature, the impedance response obtained from the macroscale tibia model constitutes an in silico prediction that still requires independent experimental validation. Fifth, model parameterization combines experimental data from bovine and human bone because of the limited availability of data simultaneously relating BV/TV, free water content, and electrical conductivity within the same species. Therefore, quantitative comparisons between these data sources should be interpreted with caution. Future studies should incorporate both conductivity and permittivity over a broader frequency range to determine whether dielectric properties or phase-related parameters provide greater sensitivity to early-stage bone deterioration than impedance magnitude alone.

\section{Conclusions}

In this work, we developed a multiscale approach based on finite element simulations and Bruggeman mixing rules to relate bone composition and microstructure to its effective conductivity at 100 kHz. For trabecular bone, the model captured the dependence of conductivity on BV/TV and free water content, accounting for their coupled variation under the adopted parameterization. For cortical bone, the anisotropic formulation incorporated the directional organization of Haversian and Volkmann canals through depolarization tensors and reproduced the behaviour observed in the FEM simulations, with higher conductivity in the axial direction.
Incorporating the effective properties into the macroscale tibia model allowed us to evaluate the electrical response associated with combined changes in bone volume fraction and cortical thickness. The scenario with lower trabecular and cortical BV/TV and reduced cortical thickness exhibited an approximately 39\% reduction in impedance relative to the reference model. Differences between electrode configurations were 2.0\% in the reference model and 0.9\% in the modified model, indicating a limited dependence of impedance on the two configurations evaluated, despite the anisotropy incorporated into the cortical tissue. The use of homogenization equations (mixing rules) provided a strategy that allowed us to incorporate the effects of composition and selected microstructural features on tissue electrical properties into macroscale simulations, without explicitly representing the entire bone microstructure.
\section*{Acknowledgments}
Argentinean National Agency for the Promotion of Science and Technology ANPCyT (PICT 2020-00457), Universidad Tecnológica Nacional (PID Grant No. MAECLP0009851TC).

\section*{Data Availability Statement}
The data that support the findings of this study, including the dataset compiled from the literature review and the numerical results from approximately 1,500 simulation runs used for model sensitivity and mesh tuning, are openly available in the repository \url{https://github.com/MEL-IFLYSIB/bone-electrical-conductivity} \cite{soft_cervantes_2026}.

\bibliographystyle{unsrt}
\bibliography{references}

\appendix

\label{sec:appendixI}
\section{Orthotropic electrical conductivity}

The electrical conductivity of an orthotropic material can be represented by a
second-order conductivity tensor. In the local coordinate system defined by
the three principal directions of electrical conduction, the conductivity
tensor is diagonal:

\begin{equation}
\boldsymbol{\sigma}_{\mathrm{local}} =
\begin{bmatrix}
\sigma_1 & 0 & 0 \\
0 & \sigma_2 & 0 \\
0 & 0 & \sigma_3
\end{bmatrix}
\end{equation}

where $\sigma_1$, $\sigma_2$, and $\sigma_3$ are the electrical conductivities
along the three principal directions of orthotropy.

For example, in cortical bone, the principal directions can be defined as
the longitudinal, circumferential, and radial directions. In this case,

\begin{equation}
\boldsymbol{\sigma}_{\mathrm{local}} =
\begin{bmatrix}
\sigma_{\mathrm{circ}} & 0 & 0 \\
0 & \sigma_{\mathrm{rad}} & 0 \\
0 & 0 & \sigma_{\mathrm{long}}
\end{bmatrix}.
\end{equation}

If the principal material directions do not coincide with the global
coordinate system, the conductivity tensor must be transformed into the
global coordinate system. This transformation is given by

\begin{equation}
\boldsymbol{\sigma}_{\mathrm{global}}
=
\mathbf{R}
\boldsymbol{\sigma}_{\mathrm{local}}
\mathbf{R}^{T}
\end{equation}

where $\mathbf{R}$ is the rotation matrix whose columns contain the three
unit vectors defining the local principal directions:

\begin{equation}
\mathbf{R}
=
\begin{bmatrix}
| & | & | \\
\mathbf{e}_1 & \mathbf{e}_2 & \mathbf{e}_3 \\
| & | & |
\end{bmatrix}.
\end{equation}

\subsection{Transversely isotropic case}

A particular case of orthotropy occurs when the electrical conductivity is
the same in two directions, i.e.,

\begin{equation}
\sigma_1 = \sigma_2 = \sigma_{\perp},
\qquad
\sigma_3 = \sigma_{\parallel}.
\end{equation}

In this case, the conductivity tensor can be expressed directly in terms of
the unit vector $\mathbf{e}_3$ defining the principal direction:

\begin{equation}
\boldsymbol{\sigma}
=
\sigma_{\perp}\mathbf{I}
+
\left(
\sigma_{\parallel}
-
\sigma_{\perp}
\right)
\mathbf{e}_3\mathbf{e}_3^{T}
\end{equation}

where $\mathbf{I}$ is the identity tensor, $\sigma_{\parallel}$ is the
conductivity along the principal direction $\mathbf{e}_3$, and
$\sigma_{\perp}$ is the conductivity in the plane perpendicular to
$\mathbf{e}_3$.

This formulation is particularly convenient when the principal direction
varies spatially, since the conductivity tensor can be defined as a function
of position,

\begin{equation}
\boldsymbol{\sigma} = \boldsymbol{\sigma}(\mathbf{x}),
\end{equation}

through the spatially varying material direction
$\mathbf{e}_3(\mathbf{x})$.

\end{document}